\documentclass{amsart}
\usepackage{amsmath, amsthm, amssymb}
\usepackage{mathrsfs}
\usepackage{mathtools} 
\usepackage{hyperref} 
\usepackage{physics} 
\usepackage{enumitem} 
\usepackage{chngcntr} 
\usepackage{a4wide} 
\usepackage[tableposition=above]{caption} 
\usepackage{graphics}
\graphicspath{{./images/}}
\usepackage{subcaption}
\usepackage{caption}
\usepackage[backend=biber,sorting=nty, style = numeric]{biblatex}
\newcounter{theorem}
\newtheorem{thm}[theorem]{Theorem}
\newtheorem{lemma}[theorem]{Lemma}
\newtheorem{prop}[theorem]{Proposition}
\newtheorem{cor}[theorem]{Corollary}
\theoremstyle{definition}
\newtheorem{defn}[theorem]{Definition}

\newtheorem{rmk/}[theorem]{Remark}
\newtheorem*{notation}{Notation}

\counterwithout{table}{section}
\counterwithin{equation}{section}
\counterwithin{theorem}{section}
\newenvironment{rmk}
	{%
  	\pushQED{\qed}\begin{rmk/}} {\popQED\end{rmk/}}

\newcommand{\vio}{\mathrm{VI}_0}

\newcommand{\scr}[1]{\mathcal{#1}}
\newcommand{\goth}[1]{\mathfrak{#1}}

\newcommand{\bb}[1]{\mathbb{#1}}
\newcommand{\conj}[1]{\overline{#1}}

\newcommand{\vp}{\varphi}
\newcommand{\ve}{\varepsilon}

\DeclareMathOperator{\ad}{ad}

\title{On Bianchi type VI$_0$ spacetimes with orthogonal perfect fluid matter}
\author{Hans Oude Groeniger}
\address{Department of Mathematics, KTH, 100 44 Stockholm, Sweden}
\email{joog@kth.se}

\begin{document}
\begin{abstract}
We study the asymptotic behaviour of Bianchi type VI$_0$
    spacetimes with orthogonal perfect fluid matter satisfying Einstein's equations.
In particular, we prove a conjecture due to Wainwright about the initial singularity of such spacetimes.
Using the expansion-normalized variables of Wainwright-Hsu, 
we demonstrate that for a generic solution the initial singularity is vacuum dominated, anisotropic and silent.
In addition, by employing known results on Bianchi backgrounds,
we obtain convergence results on the asymptotics of solutions to the Klein-Gordon equation
on all backgrounds of this type, except for one specific case.
\end{abstract}
\maketitle
%
%
\section{Introduction}
\label{sec:int}
\noindent The subject of this study is the asymptotic behaviour
    of a certain class of spatially homogeneous cosmological models.
Our interest is twofold.
First, we prove a conjecture, made over twenty years ago in \cite{DynSysCosmology},
regarding the initial singularity of Bianchi type $\vio$ spacetimes with orthogonal perfect fluid (OPF) matter
which satisfy Einstein's equations.
Resolving this conjecture allows for the study of more complicated cosmological models,
for which those of the type above appear as limit cases.
Second, by combining the asymptotic behaviour that we find with known results on Bianchi models from \cite{RingstromKleinGordon},
we are able to obtain results regarding the asymptotics of solutions to the Klein-Gordon equation on this type of model.
This fills a gap in the unified treatment of the Klein-Gordon equation on Bianchi backgrounds of \cite{RingstromKleinGordon}.

In order to reduce the complexity of Einstein's equations one often demands a high degree of symmetry 
to be present in the spacetime as well as in the matter model.
Typically, this comes in the form of isotropy or spatial homogeneity.
As a prime example, the Friedmann-Lema{\^i}tre-Robertson-Walker (FLRW) spacetimes
are both isotropic and spatially homogeneous.
Dropping the requirement of isotropy, but retaining that of spatial homogeneity,
one finds the Kantowski-Sachs spacetimes and the Bianchi spacetimes.
For a Bianchi spacetime, there is a three-dimensional group of isometries acting transitively and freely on spacelike hypersurfaces,
while for the Kantowski-Sachs spacetimes the situation is similar but the action is not free.

The underlying Lie group of a Bianchi spacetime 
    can be classified by the structure constants of the corresponding Lie algebras,
by the so-called Bianchi-Sch{\"u}cking-Behr approach, as described in
    e.g. Section 2 of \cite{EllisMacCallumHomCos}.
According to this classification, one distinguishes the class A types I, II, VI$_0$, 
    VII$_0$, VIII and IX,
and the class B types III, IV, V, VI$_h$, VII$_h$,
for a parameter~$h \neq 0 $.
As mentioned, the subject of this study are solutions of Bianchi type VI$_0$
with perfect fluid matter with velocity vector orthogonal to the group orbits.
For solutions of type I, II and VII$_0$ a similar treatment -- on which this one is largely based -- is already available,
see Sections 8-10 of \cite{RingstromBianchiIX}.

Using the orthonormal frame formalism due to Ellis and MacCallum in \cite{EllisMacCallumHomCos},
one may write the Einstein field equations for Bianchi spacetimes as evolution equations for certain cosmological variables.
This works for a variety of matter models, cf. Section 1.1.2 of \cite{DynSysCosmology}.
Then, by normalizing by the rate of expansion of the universe,
these evolution equations transform into polynomial differential equations in
	the mean curvature, the shear, the structure constants of the group and the energy density.
This approach was introduced by Wainwright and Hsu in \cite{WainwrightHsuClassA} for all class A types,
for vacuum as well as for OPF matter.
For the class B types a similar approach exists; see \cite{RadermacherVacuum} for a recent treatment.
Their approach then allows one to study the asymptotic behaviour of these quantities by employing methods from the theory of dynamical systems.
Here we are mainly interested in the approach to the initial singularity,
which is when length scales become arbitrarily small.

The main result of this article is the proof of the aforementioned conjecture.
The conjecture can be found in Section 6.3.3 of \cite{DynSysCosmology},
and is stated as Theorem \ref{thm:asymp6} below.
Translating the statement to more physical terms:
\emph{generically, the initial singularity of a Bianchi type $\vio$ orthogonal
perfect fluid solution is vacuum dominated, anisotropic and silent.} 
The genericity here is within the set of all solutions of type $\vio$ with orthogonal perfect fluid matter.

The asymptotics described below can be used to complete studies of type~$\vio$ solutions with other matter models,
for example the case of a non-orthogonal fluid or that of the magnetic type~$\vio$ solutions with OPF matter,
see \cite{HervikTiltedVI0} and \cite{LeBlancKerrWainwrightMagneticVI0} respectively.
In such compatible treatments (i.e.\ employing an approach similar to that of Wainwright and Hsu)
we find the type~$\vio$ solutions with OPF matter as limit cases.
For the much harder case of Bianchi type~VIII, the case of Bianchi type~$\vio$ appears as a limit case as well.
Although studies of the late-time dynamics exist for Bianchi type~VIII,
see \cite{HorwoodHancockTheWainwright},
the precise nature of the initial singularity remains a mystery for the non-vacuum case.

In addition to proving the conjecture of Wainwright, we also apply the derived asymptotics to fill a gap in \cite{RingstromKleinGordon}.
In that article Ringstr{\"o}m demonstrates the convergence of time derivatives
	(in certain geometrically defined time coordinates)
of solutions to the Klein-Gordon equation on Bianchi backgrounds.
His methods apply to Bianchi~$\vio$ solutions as well,
however, due to the lack of proven asymptotics, this case was left out there.
In general, the conclusions made here are very similar to those made for Bianchi types I, II and VII$_0$
	solutions with OPF matter; the precise statement is in Theorem \ref{thm:kge} below. 
\emph{Let $u$ denote a solution to the Klein-Gordon equation
	on a generic Bianchi type~$\vio$ orthogonal perfect fluid development.
Then, towards the initial singularity, 
the time derivative $u_\tau(\cdot,\tau)$ -- with respect to the expansion-normalized time -- converges exponentially to a smooth function on the underlying group.}
\subsection{Outline}
In the rest of this section we recall some preliminaries: the Bianchi types of class~A, 
the notion of a Bianchi spacetime and the stress-energy tensor of an orthogonal perfect fluid.
We also recall the Wainwright-Hsu equations, note some properties specific to the case of Bianchi type~$\vio$,
and state Theorem \ref{thm:asymp6}.
We finish the section with recalling two important tools from the theory of dynamical systems.

In Section \ref{sec:bd} we start out by describing the invariant sets in the boundary of phase space and the dynamics therein.
There we find invariant sets of types I and II as well as the vacuum case.
We also consider the dynamics within the shear invariant set, in Section \ref{sec:ga},
which contains the global attractor $P_1^+(\mathrm{VI}_0)$.
In Section \ref{sec:reg} we briefly show the regularity of the unstable sets of the equilibria from Table~\ref{tab:eq}.
The main ingredient of the proof of Theorem~\ref{thm:asymp6} is found in Section \ref{sec:pf},
where we deal with the generic case $(c)$ of convergence to the Kasner circle.

In the second half of this paper we apply Theorem \ref{thm:asymp6}
	to the Klein-Gordon equation on Bianchi type~$\vio$ backgrounds,
based on results from \cite{RingstromKleinGordon}.
The relevant theorem is formulated in Section \ref{sec:kge},
after having introduced the necessary concepts to state it in Section \ref{sec:dev}.
We consider mainly the generic case $(c)$ of Theorem \ref{thm:asymp6}, but also briefly touch on the other two cases.
\subsection{The Bianchi types of class A}
\label{sec:bia}
A connected, three-dimensional Lie group $G$ can be classified by its Lie algebra $\goth{g}$,
based on properties of the structure constants of $\goth{g}$.

We say $G$ is of class A if the Lie algebra $\goth{g}$ is \emph{unimodular},
which means that any $X \in \goth{g}$ satisfies $\trace (\ad_X) = 0$.
Any associated element (Bianchi spacetime, initial data etc.) we also dub class A if the Lie group is of class A.
If the three-dimensional Lie group is not of class A, it is of class B.

Given a basis $e_i, i = 1,2,3,$ of $(\goth{g},[\cdot,\cdot])$,
recall that the structure constants $\gamma^k_{ij}$ are defined by
\begin{equation}
[e_i,e_j] = \gamma^k_{ij} e_k.
\end{equation}
The class A algebras are precisely those for which $\gamma^k_{ik} = 0$.
We can then define the quantities
\begin{equation}
n^{ij} := \tfrac{1}{2}\big( \epsilon^{jkl} \gamma^i_{kl} + \epsilon^{ikl} \gamma^j_{kl} \big)
\end{equation}
which form a symmetric matrix $n$.

It can be shown that by choosing an appropriate basis of $\goth{g}$
this matrix can be diagonalized, with diagonal elements $n_i, i = 1,2,3$.
The Bianchi class A Lie algebras can now be classified by the signs appearing on the diagonal of the commutator matrix,
c.f. Lemma 19.8 of \cite{RingstromCauchy} and the accompanying Table \ref{tab:types}, which we copied for convenience.
This leads to the aforementioned types, of which we focus on type~$\vio$.
In case the Lie group has an algebra of type~$\vio$,
any associated element (Bianchi spacetime, initial data etc.) is also designated type~$\vio$.
\begin{table}[h!]
\captionsetup{position=below}
\begin{center}
{\renewcommand{\arraystretch}{1.2}
\begin{tabular}{l | c | c | c}
Type & $~n_1~$ & $~n_2~$ & $~n_3~$ \\ \hline
I & 0 & 0 & 0 \\ \hline
II & + & 0 & 0 \\ \hline
VI$_0$ & 0 & + & - \\ \hline
VII$_0$ & 0 & + & + \\ \hline
VIII & - & + & + \\ \hline
IX & + & + & +
\end{tabular}}
\caption{The different Bianchi types of class A, depending on the commutator matrix}
\label{tab:types}
\end{center}
\end{table}

\noindent In the case of Bianchi type~$\vio$, we may read off from Table \ref{tab:types} that one diagonal element vanishes,
while the other two have opposite signs.
The signs in Table \ref{tab:types} are, up to permutation,
precisely the signs of the $N_i, i =1,2,3$, of Section \ref{sec:wahu} below.
\subsection{Bianchi type VI$_0$ orthogonal perfect fluid solutions}
\label{sec:bias}
The spacetimes under consideration are assumed to satisfy Einstein's equations without cosmological constant.
Thus, given our Lorentz manifold $(M,h)$, we have
\begin{equation}
\label{eq:ein}
R_{\mu \nu} - \tfrac{1}{2} S h_{\mu \nu} = T_{\mu \nu}.
\end{equation}
Here $R_{\mu \nu}$ and $S$ denote the Ricci curvature tensor and the scalar curvature of $(M,h)$ respectively.
For the energy-stress tensor $T$ we take that of an orthogonal perfect fluid with linear equation of state,
which is specified below.
We demand our Lorentz manifold to be spatially homogeneous
or, more specifically, that it is a Bianchi spacetime (cf. Definition 1 of \cite{RingstromKleinGordon}).
\begin{defn}
\label{def:bias}
A \emph{Bianchi spacetime} is a Lorentz manifold $(M,h)$ of the form $M = G \times I$,
where $G$ is a connected, three-dimensional Lie group and $I = (t_-,t_+)$ an open interval;
the metric is of the form
\begin{equation}
h = - dt \otimes dt + a_{ij}(t) x^i \otimes x^j.
\end{equation}
Here the $x^j$ are the dual basis of the basis $e_i$ of the Lie algebra $\goth{g}$ of $G$.
Moreover, we require that the functions $a_{ij}(t)$ are smooth and form a positive definite matrix $a(t)$ for every $t \in I$.
\end{defn}

\begin{rmk} 
\label{rmk:lrs}
It is important to note that there are no locally rotationally symmetric (LRS) Bianchi~$\vio$ solutions.
The LRS Bianchi solutions form a special subclass of the Bianchi solutions possessing a small degree of isotropy,
see Section 1.2.2 of \cite{DynSysCosmology}.
A necessary condition for a Bianchi spacetime to be LRS is that at least two $n_i$, $i=1,2,3$, in the table above are equal,
which cannot happen for type~VI$_0$ due to the occuring signs.
However, for Bianchi~VI$_0$ solutions there is the special subclass $n_j^j =0$,
which forms an important object in our analysis;
we encounter it as the shear invariant set in Section \ref{sec:ga}.
\end{rmk}

\noindent Now, with the metric in this form, we may write the stress-energy tensor for orthogonal perfect fluid matter as
\begin{equation}
\label{eq:EMT}
T = (\rho + p) dt \otimes dt + p h.
\end{equation}
Here the variable $\rho = \rho(t)$ is known as the \emph{energy density}
and is related to the \emph{pressure} $p$ by $p = (\gamma -1) \rho$;
in particular, we demand a linear equation of state.
The parameter $\gamma$ specifies the type~of fluid under consideration,
e.g. $\gamma = 1$ corresponds to dust and $\gamma = 4/3$ to a radiation fluid.

\begin{rmk}
\label{rmk:mat}
In this article we restrict ourselves to the range $\gamma \in (\tfrac{2}{3},2)$;
in these cases both the strong and dominant energy conditions are satisfied as long as $\rho > 0$.
This is also true for $\gamma = 2$, known as the stiff fluid case, but for Bianchi~VI$_0$ this case has been well studied already,
see Section 7 of \cite{RingstromBianchiIX} for results in the dynamical systems approach,
and Example 22 of \cite{RingstromKleinGordon} for results regarding the Klein-Gordon equation.
The vacuum case, when $\rho = 0$, has been studied extensively as well,
see \cite{HeinzleRingstromFuture} and Section 23.2 of \cite{RingstromCauchy} for results in the dynamical systems approach,
and Example 24 of \cite{RingstromKleinGordon} for results regarding the Klein-Gordon equation.
We consider the vacuum case briefly in Section \ref{sec:bd}, as it appears naturally in the analysis.
\end{rmk}
\subsection{The Wainwright-Hsu variables}
\label{sec:wahu}
In what follows, we adhere to the conventions of Ringstr{\"o}m's work on the Bianchi IX attractor, see \cite{RingstromBianchiIX}, 
although the original formulation of the evolution equations in this manner is due to Wainwright and Hsu,
see \cite{WainwrightHsuClassA} or Section 6.1 of \cite{DynSysCosmology}.
The original formulation is only slightly different, and still used in
    e.g. \cite{HorwoodHancockTheWainwright}.

We use expansion-normalized variables $N_j, j =1,2,3$, $\Sigma_\pm$ and $\Omega$,
as defined in Section 21 of \cite{RingstromBianchiIX}.
We may then restrict ourselves to an invariant set of the right type
	by giving the $N_j$ the same sign as the $n_j$ in Table \ref{tab:types} above,
up to permutation, as the $N_j$ are precisely the normalized version of the $n_j, j = 1,2,3$.
This yields a hierarchy of invariant sets,
where the invariant sets of lower Bianchi types (I, II, VI$_0$ and VII$_0$) lie in the boundary of the higher types,
cf. Figure 6.1 of \cite{DynSysCosmology}.
The evolution equations take the following form:
\begin{align}
\begin{split}
N_1' &= \big(q- 4 \Sigma_+ \big) N_1, \\[1pt]
N_2' &= \big(q+2 \Sigma_+ + 2 \sqrt{3} \Sigma_- \big) N_2, \\[1pt]
N_3' &= \big(q+2 \Sigma_+ - 2 \sqrt{3} \Sigma_- \big) N_3, \\[1pt]
\Sigma_+' &= -(2-q) \Sigma_+ - 3 S_+, \\[2pt]
\Sigma_-' &= -(2-q) \Sigma_- - 3 S_-.
\end{split}
\label{eq:bianchia}
\end{align}
Since we take matter into account this system of equations is completed by the evolution equation 
\begin{equation}
\Omega' = 2(q - q^{*} ) \Omega
\end{equation}
for the \emph{normalized matter density}.
We take $\Omega \geq 0$, due to physical requirements;
$\Omega$ is related to the energy density $\rho$ through a normalization.
The set $\{\Omega = 0\}$ is invariant with respect to the dynamics;
this set is known as the \emph{vacuum case} and its orbits as \emph{vacuum solutions}.

The constant $q^{*}$ is determined by the parameter $\gamma \in (\tfrac{2}{3}, 2)$ we encountered above.
We have
\begin{equation}
q^{*}(\gamma) := \tfrac{1}{2} (3 \gamma - 2).
\end{equation} 
In the equations of motion above, the function $q$ is given by
\begin{equation}
q := q^{*} \Omega + 2 \big( \Sigma_+^2 + \Sigma_-^2 \big),
\end{equation}
and it is known as the \emph{decelaration paramater}.

The variables are not independent but related through the \emph{(Hamiltonian) constraint} 
\begin{equation}
\Omega + \Sigma_+^2 + \Sigma_-^2 + K = 1.
\end{equation}
This equation is conserved by the equations of motion (\ref{eq:bianchia}).
In particular, the variables live on a hypersurface in $\bb{R}^6$, which is preserved under the equations of motion.
The phase space is thus five-dimensional.
In the constraint, the function $K$ is short for
\begin{equation}
K := \tfrac{3}{4} \Big[N_1^2 + N_2^2 + N_3^2 -2 \big( N_1 N_2 + N_2 N_3 + N_1 N_3 \big) \Big]
\end{equation}
and lastly we have
\begin{align}
\begin{split}
S_+ &:= \tfrac{1}{2} \Big[ \big(N_2 - N_3\big)^2 - N_1 \big( 2N_1 - N_2 - N_3\big) \Big], \\
S_- &:= \tfrac{\sqrt{3}}{2} \big(N_3 - N_2 \big) \big( N_1 - N_2 - N_3 \big).
\end{split}
\end{align}
\subsection{The Bianchi type VI$_0$ phase space}
Following Table \ref{tab:types},
for Bianchi type~VI$_0$ we look at the invariant set defined by setting $N_1~=~0, ~ N_2~>~0$ and $N_3~<~0$.
The fact that this set is invariant is immediate from the system (\ref{eq:bianchia}) due to the homogeneity of the
first three equations.
A permutation of the indices of the $N_i$ can be handled by applying certain symmetries, see Section 2 of \cite{RingstromBianchiIX},
meaning our choice is without loss of generality.

If we define
\begin{align}
N_\pm &:= \tfrac{\sqrt{3} }{2} \big( N_2 \pm N_3\big),
\end{align}
then we find that the equations of motion reduce to
\begin{align}
\begin{split}
N_+' &= (q+2 \Sigma_+) N_+ +  2 \sqrt{3}   \Sigma_- N_-, \\[1pt]
N_-' &= (q+2 \Sigma_+) N_- +  2 \sqrt{3}   \Sigma_- N_+, \\[1pt]
\Sigma_+' &= -(2-q) \Sigma_+ - 2  N_-^2 , \\[2pt]
\Sigma_-' &= -(2-q) \Sigma_- - 2 \sqrt{3} N_+ N_-, \\[2pt]
\Omega' &= 2(q - q^{*} ) \Omega.
\end{split}
\label{eq:bianchi6}
\end{align}
The constraint can be written as 
\begin{equation}
\label{eq:con}
\Omega + \Sigma_+^2 + \Sigma_-^2 + N_-^2 = 1. 
\end{equation}
We define the \emph{phase space} for this system, now a hypersurface in $\bb{R}^5$, to be the set
\begin{equation}
B_1^+(\mathrm{VI}_0) := \left\{(N_+, N_-, \Sigma_+, \Sigma_-, \Omega) \in \bb{R}^5 :
	\begin{aligned} 
		&\Omega + \Sigma_+^2 + \Sigma_-^2 + N_-^2 = 1, \\
		&N_- > \abs{N_+} , ~\Omega \geq 0 
	\end{aligned}
\right\}.
\end{equation}
\begin{notation} Given $x \in \conj{B_1^+(\mathrm{VI}_0)}$ we write $(N_+(x),...,\Omega(x))$ for $(x_1, ..., x_5)$. 
Moreover, given initial data $x \in \conj{B_1^+(\mathrm{VI}_0)}$ we may write for example
\begin{equation*}
N_+(\tau) := \big( N_+ \circ \vp^\tau \big) (x) := \big(\vp^\tau(x)\big)_1
\end{equation*}
and similarly for the other coordinates.
Here and throughout this article $\vp^\tau$ denotes the flow of the dynamical system.
\end{notation}
\begin{rmk}
\label{rmk:lim}
The phase space is bounded by the constraint,
and so the closure of the phase space -- which forms an invariant set -- is compact. 
The bounds on the variables moreover grant us growth bounds since the vector field is polynomial.
This ensures that, for any point $x$ in the closure of the phase space, the flow $\vp^\tau(x)$ is complete.

Also, recall that $\alpha$- and $\omega$-limit sets are always closed and invariant. 
Since the orbits of our dynamical system are contained in a compact invariant set,
any limit set of a point in $B_1^+(\mathrm{VI}_0)$ is non-empty and connected,
see e.g. Proposition 1.1.14 of \cite{WigginsNonlinearDynSys}.
\end{rmk}

\begin{table}[h!]
\captionsetup{position=below}
\begin{center}
{\renewcommand{\arraystretch}{1.4}
\begin{tabular}{c | c | c | c}
Symbol & $\big(N_+, N_-, \Sigma_+, \Sigma_-, \Omega \big)$ & Invariant set & $u$ \\ \hline
$F$ & $\big(0,0,0,0,1\big)$ & $B(\mathrm{I})$ & 2 \\ \hline
$P_2^+(\mathrm{II})$ & $\big(\tfrac{1}{4} p^*,\tfrac{1}{4} p^*, 
	-\tfrac{1}{8} q^{*}, - \tfrac{\sqrt{3}}{8} q^{*}, 1 - \tfrac{1}{8}q^*\big)$
				& $S_2^+(\mathrm{II})$ & 1 \\ \hline
$P_3^-(\mathrm{II})$ & $ \big(\tfrac{1}{4} p^*,-\tfrac{1}{4} p^*, 
	-\tfrac{1}{8} q^{*}, \tfrac{\sqrt{3}}{8} q^{*}, 1 -\tfrac{1}{8}q^*\big)$
			 & $S_3^-(\mathrm{II})$ & 1 \\ \hline
$P_1^+(\mathrm{VI}_0)$ & $\big(0, \tfrac{1}{2}p^*, -\tfrac{1}{2} q^{*}, 0, \tfrac{1}{2}(2 - q^*) \big)$ & $S_1^+(\mathrm{VI}_0)$ & 0
\end{tabular}}
\caption{The isolated equilibria present in the closure of phase space.}
\label{tab:eq}
\end{center}
\end{table}
\noindent The closure of the phase space contains several isolated equilibria,
of which all but one reside in the boundary.
In Sections \ref{sec:bd} and \ref{sec:ga} we discuss these and their invariant manifolds in more detail.
In Table \ref{tab:eq} we list the equilibria; this table can be found (partly) in Section 6.2.1 of \cite{DynSysCosmology}.
Also, see Figure 6.3 of \cite{DynSysCosmology} for an overview of how the equilibria lie in the $(\Sigma_+, \Sigma_-)$-plane.
Here $p^*= p^*(\gamma)$ stands for $p^* (\gamma):= \sqrt{q^*(2-q^*)} \in (0,1)$.
The column with $u$ denotes the dimension of the unstable manifold relative to the phase space; 
note that all these equilibria are hyperbolic for our range of $\gamma$ and moreover, by Proposition \ref{prop:reg},
the unstable sets are indeed unstable manifolds.
The definitions of the invariant sets denoted can be found in Sections \ref{sec:bd} and \ref{sec:ga}.

The closure of the phase space also contains a circle of equilibria, called the \emph{Kasner circle}.
The circle is the unit circle in the $(\Sigma_+, \Sigma_-)$-plane and the corresponding solutions are vacuum solutions of type~I.
These are discussed in greater detail in Section \ref{sec:bd}.
For the statement of the theorem let us note that the Kasner circle contains three special points;
after removing these we are left with three components called $\scr{K}_j, j =1,2,3$.
\begin{rmk}
\label{rmk:lin}
For the computations of the linearizations around equilibria in the closure of the phase space
one rewrites $\Omega$ using (\ref{eq:con}) and subsequently ignores the direction $\partial_\Omega$.
In particular one computes the linearizations in $\bb{R}^4$;
in essence $\partial_\Omega$ only measures the behaviour normal to the hypersurface defined by (\ref{eq:con}),
which is not relevant for the desired stability results.
\end{rmk}

\noindent Now let us state our theorem.
\begin{thm}
\label{thm:asymp6}
Let $x \in B_1^+(\mathrm{VI}_0)$, with $\Omega(x) > 0$ and $\gamma \in (\tfrac{2}{3},2)$.
Then, either
\begin{enumerate}[label=(\alph*)]
\item $x$ is the equilibrium $P_1^+(\mathrm{VI}_0)$,
\item $x$ lies in the unstable manifold of one of the equilibria $F, P_2^+(\mathrm{II})$ or $P_3^-(\mathrm{II})$, or
\item $\vp^\tau(x)$ converges to a point on $\scr{K}_1$ as $\tau \to -\infty$.
\end{enumerate}
\end{thm}
\noindent 
Let us note that the unstable manifolds of case $(b)$ are of dimension not greater than two, by Table \ref{tab:eq},
so, in particular, case $(c)$ is generic, cf. Definition \ref{def:gen} below.

\subsection{Tools from dynamical systems theory}
An important tool throughout the analysis of Sections \ref{sec:bd} to \ref{sec:pf} is the monotonicity principle.
This is due to the presence of monotone functions on large parts of the phase space of Bianchi~VI$_0$.
The statement and proof of the proposition below can be found in appendix A of \cite{LeBlancKerrWainwrightMagneticVI0}.
\begin{prop}[Monotonicity principle]
\label{prop:mp}
Let $\vp^\tau$ be a flow on $\bb{R}^n$ and $S$ an invariant set of $\vp^\tau$.
Let $Z: S \to \bb{R}$ be a $C^1$-function whose range is the interval $(a,b)$,
where $a \in \bb{R} \cup \{-\infty\}, b \in \bb{R} \cup \{\infty\}$ and $a<b$. 
If $Z$ is strictly monotonically decreasing on orbits in $S$, then for all $x \in S$ we have
\begin{align}
\begin{split}
\alpha(x) &\subseteq \big\{s \in \conj{S}\setminus S ~|~ \lim_{y \to s} Z(y) \neq a \big\}, \\[2pt]
\omega(x) &\subseteq \big\{s \in \conj{S}\setminus S~ |~ \lim_{y \to s} Z(y) \neq b \big\}. 
\end{split}
\end{align}
\end{prop}

\noindent Another tool we frequently use is Gr{\"o}nwall's lemma, see e.g. Lemma 7.1 of \cite{RingstromCauchy}.
We only state a fairly elementary version, which suffices for our purposes.
Note the reversed orientation of time.
\begin{lemma}[Gr{\"o}nwall's lemma]
\label{lem:gl}
Let $I:=(a,b)$ be in an interval with $-\infty \leq a \leq b < \infty$.
Let $u \in C(\bar{I})\cap C^1(I)$ and $\beta \in C(\bar{I})$ be non-negative functions.
\begin{enumerate}[label=(\roman*)]
\item If $u$ satisfies the estimate
\begin{equation}
u'(\tau) \geq \beta(\tau) u(\tau)
\end{equation}
for all $\tau \in I$,
then we have
\begin{equation}
u(\tau) \leq u(b) \exp \left( - \int_\tau^b \beta(s) ds \right).
\end{equation}
\item If $u$ satisfies the estimate
\begin{equation}
u'(\tau) \leq - \beta(\tau) u(\tau)
\end{equation}
for all $\tau \in I$,
then we have
\begin{equation}
u(\tau) \geq u(b) \exp \left(\int_\tau^b \beta(s) ds \right).
\end{equation}
\end{enumerate}
\end{lemma}
%
%
\section{The boundary of phase space}
\label{sec:bd}
\noindent In this section we describe the boundary of the Bianchi type~VI$_0$ phase space.
Together with the shear invariant set of the next section,
this forms the so-called `skeleton' of the phase space, and plays an important role in the asymptotics,
mostly due to its invariance and the presence of monotone functions.

In order to prove Theorem \ref{thm:asymp6},
we need information about the boundary $\partial B_1^+(\mathrm{VI}_0)$ of our phase space.
Many of the proofs below use methods similar to those in \cite{RingstromBianchiIX},
while the description of invariant sets can largely be found in Chapter 6 of \cite{DynSysCosmology}.
Some of the lemmata below are simply restated for convenience,
their proofs are largely omitted and can be found in the cited sources.

We divide the boundary into invariant sets as follows:
\begin{equation}
\partial B_1^+(\mathrm{VI}_0) = B(\mathrm{I}) \cup B^+_2 (\mathrm{II})  \cup B_3^-(\mathrm{II}) 
	\cup B_1^+(\mathrm{VI}_0) \big|_{\Omega = 0}.
\end{equation}
The first three sets on the left hand side, the cases of Bianchi type I and Bianchi type II,
are due to setting $N_-^2 = N_+^2$
	(so $N_2 N_3 = 0$);
the last set appearing in the union above is the vacuum case.
\subsection{Bianchi type I}
The invariant set $B(\mathrm{I})$ consists of those points in the extended phase space satisfying 
$ N_1 = N_2 = N_3 = 0$,
or in our setting, 
\begin{equation}
B(\mathrm{I}) := \conj{B_1^+(\mathrm{VI}_0)} \cap \{N_+ = 0, N_-=0\}.
\end{equation}
The dynamics simplify considerably.
We have
\begin{align}
\begin{split}
\Sigma_+' &= -(2-q) \Sigma_+ , \\[2pt]
\Sigma_-' &= -(2-q) \Sigma_-  \\[2pt]
\Omega' &= 2\big(q - q^{*} \big) \Omega.
\end{split}
\label{eq:bianchi1}
\end{align}
If we set $\Omega = 1$ and $\Sigma_+ = \Sigma_- = 0$, then the system is in equilibrium, 
as in that case $q =q^{*}$ by (\ref{eq:con}).
This equilibrium at $(0,0,0,0,1)$ is known as $F$.

We will see below that the rest of $B(\mathrm{I})$ forms the stable manifold of $F$.
In $B_1^+(\mathrm{VI}_0)$ there is also an unstable set (of points converging to $F$ as $\tau \to -\infty$)
which we denote by $\scr{F}_{\mathrm{VI}_0}$.
This set is in fact an embedded smooth manifold of dimension 2, as we show in Proposition \ref{prop:reg}.

\subsubsection*{The Kasner circle}
Any point on the circle $\{\Sigma_+^2 + \Sigma_-^2 = 1, \Omega = 0\}$ is also an equilibrium, as in that case $q = 2$.
This circle of equilibria is known as the \emph{Kasner circle}, denoted by $\scr{K}$.
Note that each of these equilibria corresponds to a vacuum solution,
and the orbit corresponding to a vacuum solution of type~I is always contained in $\scr{K}$.
The corresponding developments (see Section \ref{sec:cor} and Section 9.1.1 of \cite{DynSysCosmology})
	are the Kasner vacuum solutions,
discovered by Kasner back in 1925.

From the system of equations (\ref{eq:bianchi1}) we deduce that the remaining orbits project to radial lines in the 
$\Sigma_+ \Sigma_-$-plane; the solutions go from the Kasner circle towards the equilibrium $F$ as time increases.
This proves the lemma below, which is Proposition 8.1 of \cite{RingstromBianchiIX}.
\begin{lemma}
\label{lem:asymp1}
Let $x \in B(\mathrm{I}) \setminus \{F\}$ for $\gamma \in (\tfrac{2}{3},2)$. 
Then
\begin{equation}
\lim_{\tau \to -\infty} (\Sigma_+, \Sigma_-, \Omega)(\tau) = \tfrac{1}{\sqrt{\sigma_+^2 + \sigma_-^2}}(\sigma_+,\sigma_-,0)
\end{equation}
where $(\sigma_+, \sigma_-)=(\Sigma_+, \Sigma_-)(x)$.
In particular, $\alpha(x)$ consists of a single point on the Kasner circle.
\end{lemma}
\noindent The lemma above yields another necessary lemma,
as a simple corollary of Lemma 4.2 of \cite{RingstromBianchiIX}.
As noted there, the assumptions of the lemma below are never satisfied,
but we need the implication to rule out the case.
\begin{lemma}
\label{lem:fimpk}
Let $x \in B_1^+(\mathrm{VI}_0)$ for $\gamma \in (\tfrac{2}{3},2)$ and assume that $F \in \alpha(x)$,
but the solution $\vp^\tau(x)$ does not converge to $F$ as $\tau \to -\infty$. 
Then $\alpha(x)$ contains a point on the Kasner circle.
\end{lemma}

\noindent There are a few points on $\scr{K}$ of special interest, the so-called \emph{Taub points}.
Dynamically they are of interest due to their eigenvalue analysis, cf. Section 6.2.2. of \cite{DynSysCosmology}.
Unlike the other points of the Kasner circle, which only have one eigenvalue 0,
the Taub points have two vanishing eigenvalues.
They are also of interest due to the geometry of the corresponding developments, see Section \ref{sec:cor},
for which the spacetimes are flat and non-silent.
\begin{defn}
\label{def:taub}
The Taub points $T_1, T_2, T_3 \in \scr{K}$ are those points on the Kasner circle with the respective coordinates
\begin{equation}
 \big(\Sigma_+, \Sigma_-\big) = \big(-1,0\big), \big(\tfrac{1}{2},
\tfrac{1}{2} \sqrt{3}\big), \big(\tfrac{1}{2}, -\tfrac{1}{2} \sqrt{3}\big).
\end{equation}
The Taub points separate the arcs $\scr{K}_j \subset \scr{K}$, where $\scr{K}_1$ is the arc bounded by $T_2$ and $T_3$ et cetera, 
see Figure \ref{fig:K1} below for an illustration of $\scr{K}_1$ and the Taub points in the 
$\Sigma_+ \Sigma_-$-plane.
The points anti-podal on $\scr{K}$ to the Taub points $T_j$ we call $Q_j, j=1,2,3$.
(So, in particular, $Q_j \in \scr{K}_j, j=1,2,3$).
\end{defn}
\begin{rmk}
\label{rmk:taub}
The arcs $\scr{K}_j$ can be characterized alternatively as $\scr{K}_j := \scr{K} \cap \{\beta_j < 0\}$,
where the $\beta_j$ are 
\begin{align}
\label{eq:betaj}
\begin{split}
\beta_1 &:= q - 4 \Sigma_+, \\[2pt]
\beta_2 &:= q + 2 \Sigma_+ + 2 \sqrt{3} \Sigma_-, \\[2pt]
\beta_3 &:= q + 2 \Sigma_+ - 2 \sqrt{3} \Sigma_-,
\end{split}
\end{align}
see Definition 6.1 of \cite{RingstromBianchiIX}.
Note that $q|_{\scr{K}} \equiv 2$, by the constraint.

\begin{figure}[b]
     \centering
         \includegraphics[width=0.45\textwidth]{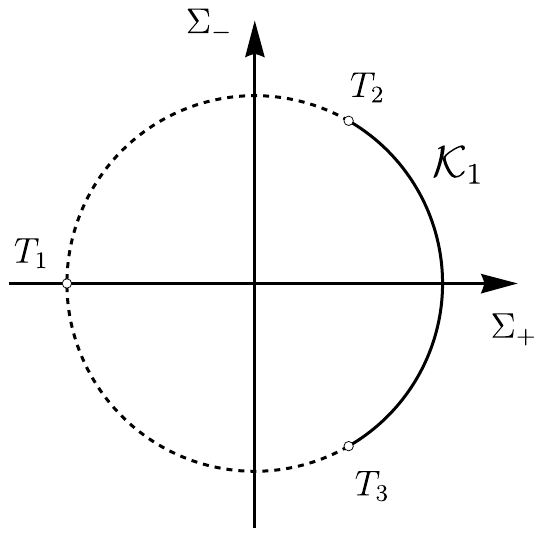}
         \caption{ The arc $\scr{K}_1$ relative to the Taub points $T_i, i \in \{1,2,3\}$
        as well as the rest of the Kanser circle (dashed).}
        \label{fig:K1}
\end{figure}

In particular, we have exponential decay of $N_j$ near the corresponding arc $\scr{K}_j$ for $j =1,2,3$,
since the $\beta_j$ are exactly $N_j'/N_j$.
The quantities $\beta_j$ are strictly positive on the arcs $\scr{K}_i$ if $i \neq j$ and we also have $\beta_j(T_j) > 0, j =1,2,3$.
However, the other Taub points are exactly where the $\beta_j$ vanish, i.e.\ $\beta_j(T_i) = 0$ if $i \neq j$.
\end{rmk}

\begin{rmk}
\label{rmk:claim}
In the introduction we claimed that, generically,
the initial singularity of a Bianchi type~VI$_0$ spacetime with OPF matter is vacuum dominated, anisotropic and silent.
The generic case $(c)$ of Theorem \ref{thm:asymp6} is convergence to $\scr{K}_1$ and
any non-special point on the Kasner circle indeed satisfies the properties mentioned.
The Taub points are the only points on $\scr{K}$ that are non-silent.

Regarding anisotropy we should note that both the Taub points $T_j$ as well as their anti-podal points $Q_j, j = 1,2,3$, are 
locally rotationally symmetric and thus contain a small degree of isotropy.
Since $\scr{K}_1$ contains the point $Q_1$, this can occur as a limit also in the generic case $(c)$ of Theorem \ref{thm:asymp6}.
However, we suspect that the only points in $B_1^+(\mathrm{VI}_0)$ that converge to $Q_1$ towards the past
are contained in the shear invariant set $S_1^+(\mathrm{VI}_0)$, which is defined below.
\end{rmk}

\noindent We note that convergence (as time goes towards the past) to the Taub points does not happen in Bianchi~VI$_0$,
as an immediate consequence of Proposition 3.1 of \cite{RingstromBianchiIX}.
We only give the necessary statement for Bianchi~VI$_0$,
which has no locally rotationally symmetric invariant set,
see Remark \ref{rmk:lrs} above.
\begin{cor}
\label{cor:not1}
For $\gamma \in (\tfrac{2}{3},2)$ there are no $x \in B_1^+(\mathrm{VI}_0)$ such that
the solution $\vp^\tau(x)$ converges to a Taub point $T_i, i=1,2,3$, as $\tau \to -\infty$.
\end{cor}

\noindent Similar to Lemma \ref{lem:fimpk}, we have a lemma concerning the Taub point $T_1$.
The proof is an adaptation from the proof of Lemma 4.1 of \cite{RingstromBianchiIX}
\begin{lemma}
\label{lem:t1impk}
Let $x \in B_1^+(\mathrm{VI}_0)$ for $\gamma \in (\tfrac{2}{3},2)$ and assume that $T_1 \in \alpha(x)$.
Then there also exists a different $\alpha$-limit point $y \in \alpha(x) \setminus \{T_1\}$ such that $\Omega(y) = 0$.
\end{lemma}
\begin{proof}
By Corollary \ref{cor:not1} we know that the solution cannot converge to the Taub point $T_1$.
Therefore there exists a $\delta > 0$ such that for any time $\tau$ there is a time $s \leq \tau$ such that $\vp^s(x) \notin B_\delta(T_1)$.
We may choose $\delta$ small enough such that for some $\ve > 0$ we have
\begin{equation}
q(x) - q^* >  \ve ~~ \text{for}~ x \in B_\ve(T_1)
\end{equation}
as $ q(T_1) = 2 > q^{*}$.

Let $(\tau_k)_{k \in \bb{N}}$ be a sequence such that $\vp^{\tau_k}(x) \to T_1$.
We may assume that $\vp^{\tau_k}(x) \in B_\delta(T_1)$ for any $k \in \bb{N}$. 
Let $s_k$ be the first time prior to $\tau_k$ such that the solution lies on the boundary of the ball,
i.e.\ $\vp^{s_k}(x) \in \partial B_\delta(T_1)$ and for $r \in (s_k, \tau_k]$ we have $\vp^r(x) \in B_\delta(T_1) $.

Then by compactness, we may select a convergent subsequence $(\vp^{s_{k_j}}(x))_j$,
yielding an $\alpha$-limit point $y$ on the boundary $\partial B_\delta (T_1)$.
In the ball $B_\delta(T_1)$ we have
\begin{equation}
\Omega' = 2 (q - q^*) \Omega \geq 2 \ve \Omega.
\end{equation}
Towards the past we see that the value $\Omega$ is exponentially decaying inside the ball,
by applying Lemma \ref{lem:gl}\emph{(i)}.
Thus we compute that,
\begin{equation}
\label{eq:decay}
\Omega(y) = \lim_{j \to \infty} \Omega(s_{k_j}) \leq \lim_{j \to \infty} \Omega(\tau_{k_j}) = 0.
\end{equation}
Thus there exists $y \in \alpha(x)$ such that $d(T_1,y) = \delta > 0$ and $\Omega(y) = 0$.
\end{proof}
\subsection{Bianchi type II vacuum dynamics}
\label{sec:bianchi2}
An invariant set of Bianchi type~II consists of points satisfying $N_j \neq 0,$ (as well as the constraint etc.)
where $j$ is exactly one of $1,2,3$, and the other two $N_i$ vanish, $i \neq j$.
In the boundary of the phase space we find two of these invariant sets.
Here we only consider the case of $B_2^+(\mathrm{II})$, i.e. the case that $N_2 > 0$ and $N_1 = N_3 = 0$.
In our setting this translates to
\begin{equation}
B_2^+(\mathrm{II}) := \conj{B_1^+(\mathrm{VI}_0)} \cap \{N_- = N_+\neq 0\}.
\end{equation}
The analysis of $B_3^-(\mathrm{II})$ is analogous (but the equality $N_- = - N_+$ is satisfied instead) so we omit it.
The dynamics in the set $B_2^+(\mathrm{II})$ are given by 
\begin{align}
\begin{split}
N_\pm' &= \big(q+2 \Sigma_+ + 2 \sqrt{3} \Sigma_-) N_\pm, \\[1pt]
\Sigma_+' &= -(2-q) \Sigma_+ - 2  N_\pm^2 , \\[2pt]
\Sigma_-' &= -(2-q) \Sigma_- - 2 \sqrt{3} N_\pm^2 , \\[2pt]
\Omega' &= 2\big(q - q^{*} \big) \Omega,
\end{split}
\label{eq:bianchi2}
\end{align}
where we denote the common value of $N_+$ and $N_-$ by $N_\pm$ (which is a multiple of $N_2$).

Type II vacuum orbits are important in order to understand of the dynamics for the higher Bianchi types,
so let us elaborate a bit on these.
By the constraint, we can rewrite the equations of motion for the vacuum case to
\begin{align}
\begin{split}
\Sigma_+' &= -2  N_\pm^2 \big( 1 + \Sigma_+\big) , \\[2pt]
\Sigma_-' &= -2  N_\pm^2 \big( \sqrt{3} + \Sigma_-\big).
\label{eq:bianchi2vac}
\end{split}
\end{align}
In particular, $1 + \Sigma_+$ is a constant multiple of $\sqrt{3} + \Sigma_-$,
so the quotient 
\begin{equation}
\label{eq:quo}
f:=\frac{1+\Sigma_+}{\sqrt{3}+\Sigma_-}
\end{equation}
is conserved along the vacuum orbit.
The intersection of the plane $\{f(x) =c\}$, for some constant $c$,
with the sphere $\{\Sigma_-^2 + \Sigma_+^2 + N_\pm^2 =1\} \subset B_2^+(\mathrm{II})$ gives the vacuum orbit through $x$ and its boundary.
For a vacuum point $x \in B_2^+(\mathrm{II})$,
the $\alpha$-limit set is but a point on $\scr{K}_2$, while $\omega(x)$ is a point on the open arc $\scr{K}_1 \cup \{T_2\} \cup \scr{K}_3$,
see Proposition 5.1 of \cite{RingstromBianchiIX}.

By considering the vacuum orbits of the other type~II systems (with $N_1 \neq 0$ or $N_3 \neq 0$),
which show similar behaviour, we obtain a map $\scr{K} \to \scr{K}$, cf. Definition 5.1 of \cite{RingstromBianchiIX}.
\begin{defn}
\label{def:kas}
We define the \emph{Kasner map} $K: \scr{K} \to \scr{K}$ as follows:
for $x \in \scr{K}$, let $K(x)$ be the point we obtain as the $\alpha$-limit of the vacuum type~II orbit
for which $x$ is the $\omega$-limit. 
\end{defn}
\noindent Note that the Taub points $T_j, j=1,2,3$, are fixed points of $K$,
while their other pre-images are precisely the $Q_j$ for the same $j$.

\subsubsection*{A lemma on the Kasner map}
An important lemma in the proof of the theorem is the following variation of Proposition 3.3 of \cite{RingstromBianchiIX},
for the case of Bianchi type~VI$_0$.
Although we prove later that $B_1^+(\mathrm{VI}_0)$ has no $\alpha$-limit points on $\scr{K}_2$ or $\scr{K}_3$,
we need the implication for the proof.
\begin{lemma}
\label{lem:bkl}
Assume $x \in B_1^+(\mathrm{VI}_0)$ has an $\alpha$-limit point $y \in \scr{K}_2 \cup \scr{K}_3$. 
Then the image $K(y)$ under the Kasner map is also contained in $\alpha(x)$.
\end{lemma}
\begin{proof}
Assume without loss of generality that the limit point $y \in \alpha(x)$ lies in $\scr{K}_2$.
Due to Remark \ref{rmk:taub},
this implies that $\beta_2(y) < 0$. 
However, $\beta_3(y) > 0$, and also 
\begin{equation}
\tfrac{1}{2}(\beta_2 + \beta_3)(y) = (q+ 2 \Sigma_+)(y) > 0.
\end{equation}
Note that the only point on $\scr{K}$ where this quantity vanishes -- which is precisely $ \frac{d}{d\tau} \log \left(N_-^2 - N_+^2 \right)$ --  is the Taub point~$T_1$, as we have $\Sigma_+ = -1$ only there.

Therefore there are $\delta>0$ and $\ve > 0$ such that in the ball $B_\delta(y)$ we have
\begin{align*}
\Omega' &= 2(2 -q^{*}) \Omega \geq \ve \Omega, \\[2pt]
(N_-^2 - N_+^2)' &= (q + 2 \Sigma_+) (N_-^2 - N_+^2) \geq \ve (N_-^2 - N_+^2),
\end{align*}
while
\begin{align*}
(N_- + N_+)' &= (q + 2 \Sigma_+ + 2 \sqrt{3} \Sigma_-)(N_- + N_+) \leq -\ve (N_- + N_+).
\end{align*}
So inside the ball $B_\delta(y)$ the value of $N_- + N_+$ is increasing exponentially towards the past,
while $\Omega, N_-^2 - N_+^2$ and thus also $N_- - N_+$ are decreasing exponentially;
this follows from Lemma \ref{lem:gl} \emph{(i)} and \emph{(ii)}.

The fact that $N_- + N_+$ is increasing exponentially implies that the solution cannot remain in the ball indefinitely,
as $N_- + N_+ \leq 2 N_- \leq 2$, by (\ref{eq:con}),
so the solution cannot converge to $y$ as $\tau \to -\infty$.
Now let $\tau_k$ be a subsequence such that $\vp^{\tau_k}(x) \to y$,
and let $s_k \leq \tau_k$ be the first time prior to $\tau_k$ such that $\vp^{s_k}(x) \in \partial B_\delta(y)$.
Then for $r \in (s_k, \tau_k]$ we have $\vp^r(x) \in B_\delta(y)$.

Therefore, by compactness, we may select a convergent subsequence $(\vp^{s_{k_j}}(x))_j$
yielding an $\alpha$-limit point $z$ on $\partial B_\delta(y)$.
It follows that $\Omega(z) = 0$ and $N_-(z) = N_+(z)$ by estimates of the form (\ref{eq:decay}),
thus $z$ is a vacuum type~I or II point.

To see that $y$ and $z$ lie in the closure of the same type~II-vacuum orbit,
we compute that the function $f$,
defined by (\ref{eq:quo}),
satisfies the estimate
\begin{align*}
|f'| &= \frac{1}{(\sqrt{3} + \Sigma_-)^2} 
		\abs{ (2-q^{*})(\sqrt{3} \Sigma_+ - \Sigma_-) \Omega 
		+ 2 \sqrt{3} (1 + \Sigma_+) \big(N_-^2 - N_+ N_- \big)}, \\[2pt]
	&\leq M \Big\{\Omega + (N_-^2 - N_+^2) + (N_- - N_+)^2 \Big\},
\end{align*}
where $M= \tfrac{4}{(\sqrt{3} -1)^3}$; we make use of the bounds provided by (\ref{eq:con}).
Therefore we estimate that for $r \in (s_k, \tau_k)$ we have
\begin{equation*}
|f'(r)| \leq M \big(\Omega (\tau_k) + (N_-^2 - N_+^2) (\tau_k) + (N_- - N_+)^2 (\tau_k) \big) ~e^{- \ve (\tau_k - r)}.
\end{equation*}
In particular, we estimate that
\begin{align*} | f(\tau_k) - f(s_k)| &\leq \int_{s_k}^{\tau_k} |f'(r)| dr, \\[2pt]
	&\leq \frac{M}{\ve} \big(\Omega (\tau_k) + (N_-^2 - N_+^2)(\tau_k) + (N_- - N_+)^2(\tau_k) \big).
\end{align*}
As $\Omega(\tau_k), (N_-^2 - N_+^2) (\tau_k)$ and $(N_- - N_+)^2(\tau_k)$ all converge to $0$ as $k \to \infty$,
we thus conclude that
\begin{equation*}
\lim_{k \to \infty} f(\tau_k) =  \lim_{k \to \infty} f(s_k).
\end{equation*}
Therefore $y$ and $z$ lie in the same plane $\{x \in \bb{R}^5: f(x) = c\}$,
so $y$ is the $\omega$-limit point of the type~II vacuum orbit through $z$,
see Section \ref{sec:bianchi2}.

If the point $z$ itself lies on the Kasner circle, we are done, as $d(y,z) >0$ and $y$ and $z$ lie on the same type~II vacuum orbit.
Thus let us assume $z \notin \scr{K}$. 
Now as $\Omega(z) = 0$, by (\ref{eq:con}) it follows that $N_-(z) \neq 0$,
which then implies that $z \in B_2^+(\mathrm{II})|_{\Omega = 0}$, since it must be of type~II.
Then the $\alpha$-limit point of the orbit through $z$, which is by definition the point $K(y)$,
also lies in $\alpha(x)$, due to the $\alpha$-limit set being closed and invariant under the flow by Remark \ref{rmk:lim}.
\end{proof}

\subsection{Bianchi type II with matter}
\noindent Each of the six type~II invariant sets comes with a non-vacuum equilibrium.
For $B_2^+(\mathrm{II})$ we have the equilibrium $P_2^+(\mathrm{II})$,
as denoted in Table \ref{tab:eq}.
Observe that the equilibrium lies in the locally rotationally symmetric invariant set 
\begin{equation}
S_2^+(\mathrm{II}) := B_2^+(\mathrm{II}) \cap \{\Sigma_- = \sqrt{3} \Sigma_+\}.
\end{equation}
With Remark \ref{rmk:lin} in mind, one may compute that, seen as part of the closure of the phase space,
the matrix of the linearization of (\ref{eq:bianchi6}) near the equilibrium $P_2^+(\mathrm{II})$ has two eigenvalues with negative real part.
We thus expect the unstable set in $B_1^+(\mathrm{VI}_0)$ to be one-dimensional;
by Proposition \ref{prop:reg} it is in fact a submanifold of dimension one, actually just a single heteroclinic orbit.

\begin{notation}
The unstable set of $P_2^+(\mathrm{II})$ within $B_1^+(\mathrm{VI}_0)$ we denote by $\scr{P}^+_{\mathrm{VI}_0}$, and 
the unstable set of $P_3^-(\mathrm{II})$ by $\scr{P}^-_{\mathrm{VI}_0}$;
the union of these two sets we write as $\scr{P}_{\mathrm{VI}_0}$, in line with Definition 4.3 of \cite{RingstromBianchiIX}.
We denote the unstable set of $F$ within $B_2^+(\mathrm{II})$ by $\scr{F}^+_{\mathrm{II}}$
and the one within $B_3^-(\mathrm{II})$ by $\scr{F}^-_{\mathrm{II}}$,
and the union of these sets is denoted by $\scr{F}_{\mathrm{II}}$.
\end{notation}

\noindent We will need information about the asymptotics towards the past for orbits of type~II.
The proposition below is Proposition 9.1 in \cite{RingstromBianchiIX}, after applying the right symmetry.
\begin{prop}
\label{prop:asymp2}
Let $x \in B_2^+(\mathrm{II})$ for $\gamma \in (\tfrac{2}{3},2)$.
If $x \notin S_2^+(\mathrm{II})$, then $\alpha(x)$ consists of a single point in $\scr{K}_1 \cup \scr{K}_3$.
Else, either
\begin{enumerate}[label=(\alph*)]
\item $x$ is the equilibrium $P_2^+(\mathrm{II})$,
\item $x$ lies in the unstable set $\scr{F}^+_{\mathrm{II}}$, or
\item the solution $\vp^\tau(x)$ converges to the Taub point $T_2$ as $ \tau \to -\infty$.
\end{enumerate}
\end{prop}
\noindent Similar to Lemma \ref{lem:fimpk},
we have the following lemma regarding the equilibrium $P_2^+(\mathrm{II})$.
This is Lemma 4.1 from \cite{RingstromBianchiIX} after applying the right symmetry.
\begin{lemma}
\label{lem:p2impk}
Let $x \in B_1^+(\mathrm{VI}_0)$ for $\gamma \in (\tfrac{2}{3},2)$ and assume that $P_2^+(\mathrm{II}) \in \alpha(x)$,
but the solution $\vp^\tau(x)$ does not converge to $P_2^+(\mathrm{II})$ as $\tau \to -\infty$. 
Then $\alpha(x)$ contains a limit point of type~\emph{II} which is not $P_2^+(\mathrm{II})$.
\end{lemma}
\subsection{The vacuum case}
The last part of the boundary of the phase space is the vacuum case, which consists of the invariant set
\begin{equation}
B_1^+(\mathrm{VI}_0)\big|_{\Omega =0}:=B_1^+(\mathrm{VI}_0) \cap \{\Omega =0\}.
\end{equation}
For a detailed study of the vacuum case for Bianchi type~VI$_0$,
including asymptotic expansions with respect to the normalized time, we refer to \cite{HeinzleRingstromFuture}.
Since the vacuum case is well-known already, we only state the relevant results on the asymptotics.
\begin{lemma}
\label{lem:vac6}
Let $x \in B_1^+(\mathrm{VI}_0)|_{\Omega = 0}$.
As $\tau \to -\infty$ the solution $\vp^\tau(x)$ converges to a point in $\scr{K}_1$,
while $\omega(x) $ consists solely of $T_1$.
\end{lemma}
\noindent 
For a proof of the statement regarding the past asymptotics set we refer to Section 22.7 of \cite{RingstromCauchy},
and regarding the $\omega$-limit set to the proof of Lemma 3.1 of \cite{HeinzleRingstromFuture}.
%
%
\section{The shear invariant set}
\label{sec:ga}
\noindent The Bianchi type~VI$_0$ phase space also contains an invariant set in its interior;
this is the shear invariant set which also contains the global sink $P_1^+(\mathrm{VI}_0)$ from Table \ref{tab:eq}.
The equilibrium has only eigenvalues with negative real part (with respect to the linearization in $B_1^+(\mathrm{VI}_0)$)
and thus it is a local attractor.
The \emph{shear invariant set} is defined as
\begin{equation}
S_1^+(\mathrm{VI}_0) :=  B_1^+(\mathrm{VI}_0) \cap \{\Sigma_- = 0, N_+ =0\}.
\end{equation}
In case of Bianchi type~VI$_0$ there are no locally rotationally symmetric solutions, cf. Remark \ref{rmk:lrs},
but one can easily check that the set above is invariant under (\ref{eq:bianchi6}).
Let us analyse the asymptotics in this set.
\begin{prop}
\label{prop:shinv}
Let $x \in S_1^+(\mathrm{VI}_0) \setminus \{P_1^+(\mathrm{VI}_0)\}$ for $\gamma \in (\tfrac{2}{3},2)$.
If $x$ does not lie in $\scr{F}_{\mathrm{VI}_0}$, then $\vp^\tau(x) \to Q_1$ as $\tau \to -\infty$.
\end{prop}
\begin{proof}
Consider the function, given in the Appendix of Section 6 of \cite{DynSysCosmology},
\begin{equation}
Z_4 := \frac{\big(N_-^2 - N_+^2\big)^m \Omega^{1-m}}{(2 + q^{*}\Sigma_+)^2},
~~~\text{where}~ m: = \frac{q^{*}}{2 + q^{*}} \in (0, \tfrac{1}{2}).
\end{equation}
The function $Z_4$ satisfies
\begin{equation}
\label{eq:z4}
Z_4' = \frac{4(2-q^{*})}{2 + q^{*} \Sigma_+}
	\bigg[ \Sigma_-^2 + \frac{\big(2\Sigma_+ +  q^{*}\big)^2}{4 - (q^{*})^2} \bigg] Z_4,
\end{equation}
and is thus strictly monotonically increasing for points in the invariant set
\begin{equation*}
B_1^+(\mathrm{VI}_0)\big|_{\Omega>0} \setminus \{P_1^+(\mathrm{VI}_0)\}.
\end{equation*}
Indeed, the derivative of $Z_4$ vanishes only if $\Sigma_- = 0$
and $\Sigma_+ = - \tfrac{1}{2} q^*$.
However, if we also demand that the time derivatives $\Sigma_\pm'$ vanish -- in order to violate monotonicity -- 
then we recover the equilibrium $P_1^+(\mathrm{VI}_0)$, due to (\ref{eq:bianchi6}) and (\ref{eq:con}).

We may also apply the monotonicity principle to the set 
\begin{equation*}
S_1^+(\mathrm{VI}_0)\big|_{\Omega>0} \setminus \{P_1^+(\mathrm{VI}_0)\}
\end{equation*}
since it is invariant.
Thus by Proposition \ref{prop:mp} we may conclude that any $y \in \alpha(x)$ lies in the boundary of this set. 
We know $P_1^+(\mathrm{VI}_0) \notin \alpha(x)$,
as the equilibrium repels backward orbits.
The $\alpha$-limit set is non-emtpy by Remark \ref{rmk:lim}. 
Also, as the limit set is contained in the boundary,
any $y \in \alpha(x)$ satisfies $\Omega(y) = 0$ or $N_-(y) = 0$.
We distinguish the case that $N_-(y) > 0$ and thus $\Omega(y) = 0$ for some $y \in \alpha(x)$,
and the case that $N_-(y) = 0$ for all $y \in \alpha(x)$.

\emph{Case 1.} Assume there is $y \in \alpha(x)$ satisfying $N_-(y) > 0$ and $\Omega(y) = 0$.
By the invariance of the $\alpha$-limit set and that of the shear invariant set,
$\alpha(x)$ contains the entire vacuum orbit $S_1^+(\mathrm{VI}_0)|_{\Omega = 0}$.
(Note that this is just a single orbit.)
The closedness of the $\alpha$-limit set then implies that both $Q_1$ and $T_1$ are in $\alpha(x)$,
cf. Lemma \ref{lem:vac6}.
However, rewriting the expression
\begin{equation}
\label{eq:est}
\Sigma_+' =-2 N_-^2(1 + \Sigma_+) - (2 - q^{*}) \Omega \Sigma_+
\end{equation}
using the constraint,
we know that $\Sigma_+$ is strictly monotonically decreasing in the interval $(-\infty, s)$
	if $\Sigma_+(s) > 0$ for some time $s$. 
Such an $s$ must exist, since $Q_1 \in \alpha(x)$.
But then $T_1 \notin \alpha(x)$, as $\lim_{\tau \to -\infty} \Sigma_+(\tau)> \Sigma_+(s) > 0$,
while at $T_1$ we have $\Sigma_+ = -1$.
Therefore the case is void.

\emph{Case 2.} Now assume that for all $y \in \alpha(x)$ we have $N_-(y) = 0$, so all $\alpha$-limit points are of Bianchi type~I.
If $F \in \alpha(x)$, then as $x \notin \scr{F}_{\mathrm{VI}_0}$, 
we may assume that there is $y \in \alpha(x)$ with $\Sigma_+(y) \neq 0$, by Lemma \ref{lem:fimpk}.
If $F \notin \alpha(x)$, the previous conclusion is immediate by the constraint.

If $\Sigma_+(y) < 0$, then after applying the flow and taking the closure we find $T_1 \in \alpha(x)$.
By Lemma \ref{lem:t1impk} we know that there is another vacuum point $y\in \alpha(x)$ which is not $T_1$. 
This other point must be $Q_1$, as these are the only points in the closure of $S_1^+(\mathrm{VI}_0)$
that satisfy $N_- = \Omega = 0$.
We may again apply the argument above to obtain a contradiction.

If $0 < \Sigma_+(y) < 1$, then by invariance and closedness of $\alpha(x)$, we find that both $Q_1$ and $F$ belong to $\alpha(x)$,
by Lemma \ref{lem:asymp1} and the fact that $B(\mathrm{I})\setminus \scr{K}$ is the stable manifold of $F$.
The argument above again leads to a contradiction.

The only remaining case is that $\Sigma_+(y) = 1$, which implies that $y = Q_1$.
If we assume that $\alpha(x)$ contains any other points, by the above we would be able to obtain yet another contradiction.
\end{proof}
\noindent Using the monotone function $Z_4$, one can readily determine the $\omega$-limit set as well.
We refer to Proposition 6.4 of \cite{DynSysCosmology};
the exponential rate follows readily from the fact that all the eigenvalues have negative real part.
\begin{prop}
\label{prop:omega}
For all $x \in B_1^+(\mathrm{VI}_0)$ with $\Omega(x) > 0$ for $\gamma \in (\tfrac{2}{3},2)$, $\vp^\tau(x)$ converges exponentially to $P_1^+(\mathrm{VI}_0)$
as $\tau \to \infty$.
\end{prop}
\noindent Outside of the shear invariant set
there is the monotone function $Z_1$ introduced in the Appendix of Section 6 of \cite{WainwrightHsuClassA}.
This will allow us to find the $\alpha$-limit set of these points, similar to Lemma 10.2 of \cite{RingstromBianchiIX} for the case of Bianchi $\mathrm{VII}_0$.
The statement regarding the $\alpha$-limit set may be found in Section 3.3 of \cite{LeBlancKerrWainwrightMagneticVI0},
but we prove a slightly stronger result.
\begin{lemma}
\label{lem:n2n3}
Let $ x \in B_1^+(\mathrm{VI}_0) \setminus S_1^+(\mathrm{VI}_0)$ for $\gamma \in (\tfrac{2}{3},2)$. 
Then the solution satisfies 
\begin{equation}
\lim_{\tau \to -\infty} \big(N_-^2 - N_+^2\big) (\tau) = 0.
\end{equation}
In particular, we have
\begin{equation}
\alpha(x) \subset B(\mathrm{I}) \cup B_2^+(\mathrm{II}) \cup B_3^-(\mathrm{II}).
\end{equation}
\end{lemma}
\begin{proof}
Assume to the contrary, then by compactness of the closure of the phase space, 
we can find $y \in \alpha(x)$ such that $N_-^2(y) > N_+^2(y)$.

Consider the monotone function, as given in \cite{WainwrightHsuClassA},
\begin{equation}
\label{eq:z1}
Z_{1} := \frac{\Sigma_-^2 + N_+^2}{N_-^2 - N_+^2}.
\end{equation}
This is well-defined and positive on the invariant set $B_1^+(\mathrm{VI}_0) \setminus S_1^+(\mathrm{VI}_0)$.
It is indeed strictly monotonically decreasing; note that we have
\begin{equation}
Z_{1}' = -4 \frac{\Sigma_-^2 \big(1+\Sigma_+\big)}{\Sigma_-^2 + N_+^2} Z_{1}.
\end{equation}
Firstly, by (\ref{eq:con}), we must have $ \Sigma_+(\tau) > -1$, as $N_-(\tau) > 0$ for any $\tau \in \bb{R}$. 
Secondly, if $\Sigma_-(\tau) =0$, we must have $\Sigma_-'(\tau) \neq 0$,
else we may show that $x \in S^+_1(\mathrm{VI}_0)$ by invariance.
Then by integrating it follows that $Z_{1} (\tau_0) > Z_1(\tau_1)$ for any $\tau_0 < \tau_1$,
so the function $Z_1$ is indeed strictly decreasing.

By Proposition \ref{prop:mp} it follows that
\begin{equation*}
y \in \Big(\conj{B_1^+(\mathrm{VI}_0)} \setminus B_1^+(\mathrm{VI}_0) \Big)\cup S_1^+(\mathrm{VI}_0) 
	\subset \{ N_-^2 = N_+^2 ~\text{or}~ \Sigma_-^2 + N_+^2 = 0 \}
\end{equation*}
However, since $Z_1$ is decreasing and $Z_1(x) > 0$,
it cannot happen that $\Sigma_-^2(y) + N_+^2(y) = 0$ if not also $N_-^2(y) = N_+^2(y)$.
\end{proof}
%
%
\section{Regularity of the unstable sets of $F$, $P_2^+(\mathrm{II})$ and $P_3^-(\mathrm{II})$}
\label{sec:reg}
\noindent For the strength of Theorem \ref{thm:asymp6}, 
it would be interesting to know whether the sets of points
which converge towards the past to $F$ and $P_2^+(\mathrm{II}), P_3^-(\mathrm{II})$,
	are in fact smooth manifolds,
so as to verify that they are sets of measure zero within the phase space.
The ideas here are similar to those of Theorem 16.1 of \cite{RingstromBianchiIX},
but here we make use of the information we gathered about $B_1^+(\mathrm{VI}_0)$ in Proposition \ref{prop:omega}.
The set $\scr{F}_{\mathrm{II}}$ is already covered in \cite{RingstromBianchiIX}, so we omit it.
\begin{prop}
\label{prop:reg}
The unstable sets $\scr{P}^+_{\mathrm{VI}_0}, \scr{P}^-_{\mathrm{VI}_0}$ and $\scr{F}_{\mathrm{VI}_0}$
	are (embedded) smooth submanifolds of $\bb{R}^4$ of dimensions~$1,1$ and $2$ respectively.
\end{prop}
\begin{proof}
We deal with the case of $\scr{F}_{\mathrm{VI}_0}$, the other two cases being similar and easier to handle,
as they each consist of just a single heteroclinic orbit.\footnote{
Generally there are two one-dimensional unstable manifolds in case of a single positive eigenvalue,
but due to the restrictions of our phase space,
one of those orbits is not part of the phase space; for $F$ this comes from the restriction $N_- > 0$,
and for $P_j^\pm(\mathrm{II})$ it comes from the restriction $N_- > \abs{N_+}$.}

We claim first that there is a neighbourhood $U$ of $F$ such that $U \cap \scr{F}_{\mathrm{VI}_0} = U \cap M_F$,
where $M_F \subset B_1^+(\mathrm{VI}_0)$ is the local unstable manifold of dimension two guaranteed by the Stable Manifold Theorem
	(see e.g. Theorem 1.3.2 of \cite{GuckenheimerHolmes}).
Let us assume such a $U$ does not exist;
we may then select a sequence of points $(x_k)_k \subset \scr{F}_{\mathrm{VI}_0}$, $x_k \notin M_F$,
converging to $F$ as $k \to \infty$ and such that $\vp^\tau(x_k)$ converges to $F$ as $\tau \to -\infty$.

We may choose a neighbourhood $V$ of $F$ with smooth boundary $\partial V$,
such that $M_F$ intersects $\partial V$ transversally. 
The resulting one-dimensional submanifold, $S:=\partial V \cap M_F$, may be assumed to be closed so in particular compact.
Each point $y$ on $S$ uniquely defines a heteroclinic orbit $\scr{O}(y) \subset \scr{F}_{\mathrm{VI}_0}$,
from $F$ to $P_1^+(\mathrm{VI}_0)$, as $\Omega(y) >0, N_-(y) >0$
	and thus $\vp^\tau(y) \to P_1^+(\mathrm{VI}_0)$ by Proposition~\ref{prop:omega}.
On the other hand, each orbit in $\scr{F}_{\mathrm{VI}_0}$ contains a single point $y \in S$.

Without loss of generality we may assume that $(x_k)_k \subset V$.
Now define for each $x_k$ the time $T_k$ as the smallest positive number
	such that $\vp^{T_k}(x_k)$ lies on the boundary $\partial V$.
Also, define $y_k$ as the sole element of $ \scr{O}(x_k) \cap S$.

Choose a neighbourhood $W$ of $P_1^+(\mathrm{VI}_0)$, not intersecting $V$, homeomorphic to a ball and
such that orbits entering $W$ do not leave it.
This is possible due to the Hartman-Grobman theorem (see e.g. Theorem 1.3.1 of \cite{GuckenheimerHolmes}) and all eigenvalues of the linearization at $P_1^+(\mathrm{VI}_0)$ having negative real part.
We then assign to each $y \in S$ the time $\tau_y$ to reach $\partial W$.
This assignment is continuous -- since $(\tau, x) \mapsto \vp^\tau(x)$ is smooth -- and thus by compactness also bounded.

Then we estimate that
\begin{equation*}
T_k \leq \tau_{y_k} \leq \max_{y \in S}~ \tau_y.
\end{equation*}
In particular, the sequence $(T_k)_k$ is bounded,
and thus we may choose a convergent subsequence $T_{k_j} \to T_0$.
Now note that 
\begin{equation*}
\lim_{j \to \infty} \vp^{T_{k_j}} (x_{k_j}) = \vp^{T_0}(F) = F
\end{equation*}
as $F$ is an equilibrium.
But as $\vp^{T_{k_j}}(x_{k_j}) \subset \partial W$,
we may choose a convergent subsequence there to obtain a contradiction.

Given any $x \in \scr{F}_{\mathrm{VI}_0}$ we can find a time $s(x)$ such that $\vp^{-s(x)}(x) \in U$.
In particular, by invariance under the flow of the set $\scr{F}_{\mathrm{VI}_0}$, we find that $\vp^{-s(x)}(x) \in U \cap M_F$.
The diffeomorphisms $\vp^\tau$ thus induce the structure of a smooth submanifold on $\scr{F}_{\mathrm{VI}_0}$
from the one on $M_F$.
\end{proof}
\noindent Knowing that the unstable manifolds of $F, P_2^+(\mathrm{II})$ and $P_3^-(\mathrm{II})$ are small,
it makes sense to define the points in the phase space outside of these sets, and not being the single point $P_1^+(\mathrm{VI}_0)$,
as \emph{generic}.
We define it for the case of $B_1^+(\mathrm{VI}_0)$, but one can obtain similar definitions in the other cases
$B_j^\pm(\mathrm{VI}_0)$ by applying the appropriate symmetries.
\begin{defn}
\label{def:gen}
We call a point $x \in B_1^+(\mathrm{VI}_0)$ \emph{generic} if
	it is not the equilibrium $P_1^+(\mathrm{VI}_0)$
and it is not contained in the unstable manifold $\scr{P}_{\mathrm{VI}_0}^+$,
$\scr{P}_{\mathrm{VI}_0}^-$ or $\scr{F}_{\mathrm{VI}_0}$.
\end{defn}
%
%
\section{Convergence to $\mathcal{K}_1$ for generic points}
\label{sec:pf}
\noindent We can now study the limit behaviour for generic points outside of the shear invariant set,
applying a proof similar to that of Proposition 10.2 of \cite{RingstromBianchiIX}.
This result, along with Propositions \ref{prop:shinv} and \ref{prop:reg},
proves Theorem \ref{thm:asymp6}.
\begin{prop}
\label{prop:gen}
Let $x \in B_1^+(\mathrm{VI}_0) \setminus S_1^+(\mathrm{VI}_0)$ be generic,
for $\gamma \in (\tfrac{2}{3},2)$.
Then $\vp^\tau(x)$ converges to a point in $\scr{K}_1$.
\end{prop}
\begin{proof}
Firstly, by Remark \ref{rmk:lim} the $\alpha$-limit set $\alpha(x)$ is non-emtpy.
By Lemma \ref{lem:n2n3}, we know that 
\begin{equation}
\alpha(x) \subset B(\mathrm{I}) \cup B_2^+(\mathrm{II}) \cup B_3^-(\mathrm{II}).
\end{equation}
We begin by showing that $\alpha(x) \cap \scr{K} \neq \emptyset$. 
We distinguish the case that there is a limit point of type~I,
and the case that there is a limit point of type~II.

\emph{Case 1.} Assume there is a limit point $y \in \alpha(x)$ of type~I.
If $y \neq F$, then by Lemma \ref{lem:asymp1} we find an $\alpha$-limit point in $\scr{K}$.
But as convergence to $F$ was excluded by assuming $x \notin \scr{F}_{\mathrm{VI}_0}$,
we know that if $y = F$, then Lemma \ref{lem:fimpk} applies, so that $\alpha(x)$ contains a point in $\scr{K}$.

\emph{Case 2.} If there exists a $y \in \alpha(x)$ is of type~II, we may assume $y \notin \scr{F}_{\mathrm{II}}$,
else we are back in the case above, by invariance and closedness of $\alpha(x)$.
Moreover, we may assume that $y$ is not one of the equilibria $P_2^+(\mathrm{II})$ or $P_3^-(\mathrm{II})$,
by Lemma \ref{lem:p2impk}.
Note again that $x$ cannot converge to these equilibria, since we assumed $x \notin \scr{P}_{\mathrm{VI}_0}$.
Then, by Proposition \ref{prop:asymp2}, $y \in \scr{K}$.

Now we shall show that $\alpha(x)$ contains an element of $\conj{\scr{K}_1}$.
Let $y \in \alpha(x) \cap \scr{K}$.
In case that $y \in \scr{K}_2 \cup \scr{K}_3$,
we may apply the Kasner map a finite number of times to find an $\alpha$-limit point in $\conj{\scr{K}_1}$ by Lemma \ref{lem:bkl}.
In case that $y = T_1$,
we use Lemma \ref{lem:t1impk} to find $z \in \alpha(x) \cap \{\Omega =0\}$ which is not $T_1$.
If $z$ is of type~I, then it must lie on the Kasner circle and since it is not $T_1$ we are done by the case above.
If $z$ is of type~II, then by closure and invariance of $\alpha(x)$ 
there exists $w \in \scr{K}$ such that both $w$ and $K(w) \neq w$ are contained in $\alpha(x)$.
Since at most one of these is $T_1$, we are again done by the case above.

Next we claim that $\alpha(x)$ consists of only a single point $z \in \scr{K}_1$.
Write $N_-^2 = 1 - \Omega - \Sigma_+^2 - \Sigma_-^2$ using the constraint.
Then we obtain that
\begin{equation}
\Sigma_+' =-(2 - 2 \Omega - 2 \Sigma_+^2 - 2 \Sigma_-^2)(1 + \Sigma_+) - (2 - q^{*}) \Omega \Sigma_+,
\end{equation}
so in particular: if $\Sigma_+ > 0$,
then it is strictly monotonically decreasing and thus has a unique limit as $\tau \to -\infty$.
Moreover, $\Omega$ must converge to $0$, else $\Sigma_+$ becomes unbounded
thus violating (\ref{eq:con}).
Indeed, we have the inequality
\begin{equation}
\Sigma_+' \leq -(2-q^*) \Omega \Sigma_+
\end{equation}
and by Lemma \ref{lem:gl}\emph{(ii)}, we conclude that 
\begin{equation}
\Sigma_+(\tau) \geq \Sigma_+(\tau_0) e^{(2-q^*)\int^{\tau_0}_{\tau} \Omega(s) ds}
\end{equation}
for any $\tau < \tau_0$, where $\tau_0$ is any time such that $\Sigma_+(\tau_0) > 0$.
In particular, $\int^{\tau_0}_{\tau} \Omega(s) ds < \infty$, so $\Omega$ is integrable.
Since the function $\Omega(\tau)$ is continuous, positive and has bounded derivative
-- recall that the closure of phase space is compact and the vector field polynomial --
we thus have $\lim_{\tau \to- \infty} \Omega(\tau) = 0$.
In the same way $\Sigma_+^2 + \Sigma_-^2$ converges to $1$.
Then, by the constraint, also $N_-$ and thus also $N_+$ converge to $0$.
The limit of all coordinates is thus unique, up to the sign of $\Sigma_-$. 
However, by connectedness of $\alpha(x)$, we conclude that this sign is uniquely determined.
Therefore $\vp^\tau(x)$ converges to a single point in $\conj{\scr{K}}_1$ as $\tau \to -\infty$. 
We know from Corollary \ref{cor:not1} that $x \in B_1^+(\mathrm{VI}_0)$ cannot converge to $T_2$ or $T_3$,
so we conclude that $z \in \scr{K}_1$.
\end{proof}
\noindent For the subsequent sections, the following will be of use.
\begin{cor}
\label{cor:q}
Let $x \in B_1^+(\mathrm{VI}_0)$ be generic, for $\gamma \in (\tfrac{2}{3},2)$.
Then $2 - q(\tau)$ is positive and it converges to $0$ exponentially as $\tau \to -\infty$.
\end{cor}
\begin{proof}
We know that for a generic $x \in B_1^+(\mathrm{VI}_0)$ the solution $\vp^\tau(x)$ converges to a point in $\scr{K}_1$,
by Propositions \ref{prop:shinv} and \ref{prop:gen}.
In particular, $q(\tau) \to 2$ as $\tau \to -\infty$, so we only have to show the positivity and the exponential rate.

The exponential rate follows from the exponential rates at which $N_-$ and $\Omega$ converge to 0,
combined with the constraint.
We have
\begin{equation}
N_-' \geq \big(q + 2 \Sigma_+ - 2 \sqrt{3} \abs{\Sigma_-}\big) N_-.
\end{equation}
Since the factor in front of $N_-$ above is positive everywhere on $\scr{K}_1$, by Remark \ref{rmk:taub},
there are $s = s(x,\gamma)\in \bb{R}$ and $\eta = \eta(x,\gamma) \in (0, 2 - q^*)$
such that for any $\tau \leq s$ we have 
\begin{equation}
N_-(\tau) \leq N_-(s) e^{-\eta(s - \tau)}.
\end{equation}
This follows from Gr{\"o}nwall's lemma.

Regarding $\Omega$, we have 
\begin{equation}
\Omega' = 2 (q - q^*) \Omega.
\end{equation}
Thus select a time $s$ -- which is possibly different from the one above,
but take $s$ to be the smaller one between them --
such that for $\tau \leq s$ we have $q(\tau) - q^* > \eta$.
This is possible as $q \to 2$ and $q^* \in (0,2)$ for $\gamma \in (\tfrac{2}{3},2)$.
Then it is clear that also
\begin{equation}
\Omega(\tau) \leq \Omega(s) e^{-2 \eta (s- \tau)}.
\end{equation}
By the constraint equation \ref{eq:con}, we can write
\begin{equation}
2 - q(\tau) = (2 - q^*) \Omega(\tau) + 2 N_-(\tau)^2
\end{equation}
The positivity follows now from the function $N_-$ not vanishing in the phase space.
By the equation above, we find that, for $\tau \leq s$, there must be a constant $C = C(x, \gamma) >0$ such that
\begin{equation}
2 - q(\tau) \leq C e^{2 \eta \tau }.
\end{equation}
Note that $C$ and $\eta$ are independent of $\tau$, they only depend on $\gamma$ and the initial data $x$.
\end{proof} 
%
%
\section{Bianchi type VI$_0$ developments}
\label{sec:dev}
\noindent In this section we introduce some concepts
	which are necessary to translate Theorem \ref{thm:asymp6} to the geometrical setting.
This allows us to apply the theorem to the Klein-Gordon equation, 
by making use of the machinery developed in \cite{RingstromKleinGordon}, an article on asymptotics of solutions to the Klein-Gordon equation on Bianchi backgrounds.

The material of this section is not new;
we refer the reader to \cite{RingstromKleinGordon}, \cite{RingstromBianchiIX} and \cite{WainwrightHsuClassA} in particular.
We refrain from going into details too much
	and restrict ourselves to introducing the concepts necessary to state and prove Theorem \ref{thm:kge}.

\subsection{Orthogonal perfect fluid data}
Recalling Definition \ref{def:bias},
consider the induced metric $g_t~=~h|_{TG_t \times TG_t}$ on $G_t: = \{t\} \times G$.
As $(G_t ,g_t)$ is a spacelike submanifold, orthogonal to $\partial_t$, 
we have, dropping the $t$-dependencies,
\begin{equation}
\label{eq:data1}
S - k^{ij} k_{ij} + (\tr_g k)^2 = 2 \rho,
\end{equation}
as well as
\begin{equation}
\label{eq:data2}
D_{i} (\tr_g k) - D_{j} \bar{k}^i_j =0.
\end{equation}
These are the Hamiltonian and momentum constraint equations
    for an orthogonal perfect fluid, respectively.
Here $k$, $S$ and $D$ are respectively the second fundamental form, 
scalar curvature and Levi-Civita connection of $(\{t\} \times G,g_t)$.

Taking a different perspective, we may also specify the initial data to satisfy these equations, cf. Definition 9 of \cite{RingstromKleinGordon}.
\begin{defn}
\label{def:bopfd}
\emph{Bianchi orthogonal perfect fluid data} $(G, g, k, \rho_0)$ consist of a connected three-dimensional Lie group $G$,
a left invariant metric $g$ on $G$,
a left invariant, symmetric, covariant two-tensor field $k$ on $G$,
and a non-negative constant $\rho_0$, 
satisfying equations (\ref{eq:data1}) and (\ref{eq:data2}).
\end{defn}

\noindent Fluid data of class A, in the sense above, each give a point in the phase space of the system (\ref{eq:bianchia}).
For this we refer to Lemma 21.2 of \cite{RingstromBianchiIX} and Section 17.1 of \cite{RingstromKleinGordon}.
By permutation we may assume that Bianchi~VI$_0$ orthogonal fluid data correspond to a point in $B_1^+(\mathrm{VI}_0)$
without loss of generality.
Conversely, see p. 1419 of \cite{WainwrightHsuClassA},
the orbit of a point in the phase space of (\ref{eq:bianchia}) corresponds to 
	a one-parameter family of conformally-related developments.
This is due to an arbitrariness in determining the length scale in the cosmological model.

\begin{defn}
\label{def:data}
Given Bianchi~VI$_0$ orthogonal perfect fluid data,
we call it \emph{generic type $\mathrm{VI}_0$ data}
if the corresponding point $x \in B_1^{+} (\mathrm{VI}_0)$ is generic in the sense of Definition \ref{def:gen}.
\end{defn}

\subsection{The initial singularity}
We are interested in Bianchi spacetimes with a certain singularity towards the past.
The type of singularity is defined precisely in Definition 2 of \cite{RingstromKleinGordon}, which we copy for convenience.
\begin{defn}
\label{def:mvs}
Let $(M,h)$ be a Bianchi spacetime.
Then $t_-$ is a \emph{monotone volume singularity} if there exists a time $t_0 \in I$
	such that the mean curvature $\theta(t)$ of the hypersurface $G \times \{t\}$ is strictly positive on the interval $(t_-,t_0)$,
and if the function $\tau$, defined as
\begin{equation}
\tau(t):= \tfrac{1}{3} \log \sqrt{ \det a(t)}
\end{equation}
satisfies $\tau(t) \to -\infty$ as $t \to t_-$.
\end{defn}
\begin{rmk}
The function $\tau$ is known as the \emph{logarithmic volume density},
and satisfies $\tau'(t) = \theta/3$.
In what follows we employ $\tau$ as a new time-coordinate.
However, it is also precisely the time coordinate used in the Wainwright-Hsu equations.

In Theorem \ref{thm:kge} we also encounter the time coordinate $\sigma$,
defined implicitly by
\begin{equation}
\frac{d \sigma}{dt} = \frac{1}{3} \frac{1}{\sqrt{ \det a(t)}},
\end{equation}
together with the condition that $\sigma(t) = 0$ for the same time $t$ for which $\tau(t) = 0$.
We note that if the Lie group is of type~VI$_0$ then $\sigma \to -\infty$ corresponds to $\tau \to -\infty$, 
see Lemma 52 of \cite{RingstromKleinGordon}.
\end{rmk}

\begin{rmk}
\label{rmk:dec}
In the present context, one may write the Raychaudhuri equation as
\begin{equation}
\theta'(\tau) = -(1 + q) \theta (\tau)
\end{equation}
where $q$ is the \emph{deceleration parameter},
which coincides with the function $q(\tau)$ introduced in Section \ref{sec:wahu}.
Recall that $2 - q(\tau)$ is positive and converges exponentially to $0$ for generic $x \in B_1^+(\mathrm{VI}_0)$, by Corollary \ref{cor:q}.
In particular, $2 - q(\tau)$ is an integrable function.
It follows that for generic type~$\vio$ data we have
\begin{equation}
\theta(\tau) \geq C e^{-3\tau},
\end{equation}
where $\tau <0$ and 
\begin{equation}
C = \theta(0) \exp\left(\int_{-\infty}^0 2-q(\tau') d\tau' \right) < \infty.
\end{equation}
Note that $C$ depends on both $\gamma$ and the data.
\end{rmk}
\subsection{Class A developments}
\label{sec:cor}
Given Bianchi class A orthogonal perfect fluid data $(G,g,k, \rho_0)$,
there is a corresponding development, which is in fact a Bianchi spacetime.
This is the content of Lemma 21.2 of \cite{RingstromBianchiIX};
this development is called the \emph{corresponding class A development}.

\begin{rmk}
\label{rmk:max}
In case that the underlying Lie group is of type~$\vio$, the solution always has a monotone volume singularity. 
Indeed, for Bianchi~$\vio$, the lack of locally rotationally symmetric developments, see Remark \ref{rmk:lrs}, implies
that all of the corresponding developments are past causally geodesically incomplete and future causally geodescially complete,
by Lemmas 21.5 and 21.8 of \cite{RingstromBianchiIX}.
By Lemma 21.8 of \cite{RingstromBianchiIX}, for type~$\vio$ the interval $I$ is of the form $I = (t_-, \infty)$, with $t_- > -\infty$. 
In the new time coordinate $\tau$ this interval gets mapped to $\bb{R}$, by Lemma 22.4 of \cite{RingstromBianchiIX},
so we always have a monotone volume singularity for Bianchi~type~$\vio$ with OPF matter.
\end{rmk}

\begin{rmk}
\label{rmk:curv}
Let us note that for a class A development $(M,h)$ not of type~IX, the scalar curvature $S$ of constant$-t$ hypersurfaces $(M_t,g_t)$ is nonpositive;
for those Bianchi types the function $\gamma_S := \max\{0,S\}/\theta^2$ introduced in Theorem 7 of \cite{RingstromKleinGordon} is identically zero,
see the remarks after Remark 8 of \cite{RingstromKleinGordon}.
\end{rmk}
%
%
\section{Klein-Gordon equation on Bianchi type VI$_0$ backgrounds.}
\label{sec:kge}
\noindent We now tackle the Klein-Gordon equation on type~$\vio$ backgrounds,
by combining results from \cite{RingstromKleinGordon} with Theorem \ref{thm:asymp6} above.
This fills a gap in the results of \cite{RingstromKleinGordon}, specifically in Example 29 and Section 15.2,
which was due to lack of proven asymptotics for type~$\vio$ in the Wainwright-Hsu variables.

Given a Lorentz manifold $(M,h)$, the Klein-Gordon equation reads
\begin{equation}
\label{eq:kg}
\Box_h u = m^2 u
\end{equation}
where $m \in \bb{R}$ is a constant.
Now we are in a position to state our theorem.
In the statement below $\langle \sigma \rangle : = \sqrt{ 1 + \sigma^2}$ denotes the Japanese bracket.
\begin{thm}
\label{thm:kge}
Let $(G, g, k, \rho_0)$ be Bianchi orthogonal perfect fluid data for $\gamma \in (\tfrac{2}{3},2)$,
with $G$ of type $\vio$ and with $\rho_0 > 0$. 
Assume that the data are generic, and denote the corresponding development by $(M,h)$.

Then there exist constants $ \lambda, \nu> 0$,
such that for any smooth solution $u \in C^\infty(M)$ of the Klein-Gordon equation (\ref{eq:kg}) on $(M,h)$,
there are smooth functions $ v, w, \phi, \psi \in C^\infty(G)$,
such that for each compact subset $K \subseteq G$ and for each $l \in \bb{N}_0$,
we have
\begin{equation}
\norm{u_\tau(\cdot,\tau) - v}_{C^l(K)} + \norm{u (\cdot,\tau) - \tau \cdot v  - w}_{C^l(K)}
	\leq C_{K,l} \langle \tau \rangle e^{\lambda \tau}
\end{equation}
for all $\tau \leq 0$, as well as
\begin{equation}
\norm{u_\sigma(\cdot,\sigma) - \phi}_{C^l(K)} + \norm{u (\cdot,\sigma) - \sigma \cdot \phi  - \psi}_{C^l(K)}
	\leq C_{K,l} \langle \sigma \rangle e^{\nu \sigma}
\end{equation}
for all $\sigma \leq 0$. 
\end{thm}
\begin{proof}
We wish to apply Proposition 19 of \cite{RingstromKleinGordon}.
From Sections \ref{sec:bias} and \ref{sec:dev}, it is clear that the following is satisfied:
$(M,h)$ has a monotone volume singularity $t_-$, by Remark \ref{rmk:max};
the metric $h$ solves Einstein's equation (\ref{eq:ein});
$\rho \geq p$, as well as $\rho \geq 0$ and $\Lambda \geq 0$, by Remark \ref{rmk:mat},
where we keep in mind that if $\rho_0 > 0$ then $\rho > 0$ throughout the development.
Moreover, for type $\vio$ spacetimes, the scalar curvature $S$ of constant$-t$ hypersurfaces $(M_t,g_t)$ is nonpositive, 
implying that $\gamma_S \equiv 0$, by Remark \ref{rmk:curv}.

Note that the function $\vp_0$ is simply $-m^2$ in our setting, 
and thus, by Remark \ref{rmk:dec},
\begin{equation}
\abs{\hat{\vp}_0 (\tau)} = 9 \theta^{-2} \abs{\vp_0(\tau)} \leq 9 C^{-2} m^2 \exp(6 \tau).
\end{equation}
where $C>0$ is a constant depending on the value of $\gamma$ and the initial data.
Moreover, we know that for generic data $2 - q(\tau)$ decays exponentially, by Corollary \ref{cor:q}.
Lastly, we know that $\norm{\hat{a}^{-1}(\tau)}$ decays exponentially.
Indeed, for generic data as in Definition \ref{def:data},
the corresponding backward orbit of (\ref{eq:bianchia}) converges to a point on one of the Kasner arcs $K_j, j=1,2,3$,
by Theorem \ref{thm:asymp6}.
From Section 17.1 of \cite{RingstromKleinGordon}, we gather that this implies the desired estimate for $\norm{\hat{a}^{-1}(\tau)}$.

We have satisfied all assumptions of Proposition 19 of \cite{RingstromKleinGordon},
establishing the proof.
\end{proof}
\subsection{Non-generic data}
For developments corresponding to case $(b)$ of Theorem \ref{thm:asymp6},
asymptotics to solutions of the Klein-Gordon equation are discussed in Example 35 of \cite{RingstromKleinGordon}.
Therefore, only case $(a)$ remains. 

For the development corresponding to the equilibrium $P_1^+(\mathrm{VI}_0)$ we may compute that
\begin{equation}
\hat{a}^{11}(\tau) = C_{a,1} |N_2 N_3|(\tau) = \tfrac{1}{12} C_{a,1} (p^*)^2,
\end{equation}
where $C_{a,1} >0$ is a constant, cf. Section 14.1 of \cite{RingstromKleinGordon}.
Since $p^* \in (0,1)$ for $\gamma \in (\tfrac{2}{3},2)$, 
we find that the norm $\norm{\hat{a}^{-1}}$ remains bounded away from zero.
Thus we know that
\begin{equation}
\norm{\hat{a}^{-1}}^{1/2} \notin L^1\big((-\infty,0]\big),
\end{equation}
meaning the singularity is not silent.
Therefore we cannot apply Proposition 19 of \cite{RingstromKleinGordon} to find the desired asymptotics.

\section*{Acknowledgements}
\noindent The author sincerely thanks Hans Ringstr{\"o}m for his many comments and suggestions,
the anonymous reviewer for their careful reading of the manuscript,
and Bernhard Brehm for his lecture series on the dynamical systems approach.
This research was funded by the Swedish research council,
dnr. 2017-03863.

\printbibliography
\end{document}